\documentstyle[pra,aps,twocolumn,psfig]{revtex}
\draft
\begin{document}

\title{Maximum-likelihood estimation of quantum measurement}

\author{Jarom\'{\i}r Fiur\'{a}\v{s}ek}

\address{Department of Optics, Palack\'{y} University, 17. listopadu 50,
772 07 Olomouc, Czech Republic}

\maketitle

\begin{abstract}
 Maximum likelihood estimation is applied to  the
determination of an unknown quantum measurement.
The measuring apparatus performs  measurements on many
different quantum states and
the positive operator-valued measures governing the measurement
statistics are  then inferred from the
collected data via Maximum-likelihood principle. In contrast to
the procedures based on linear inversion, our
approach always provides physically sensible result. We illustrate
the method on the case of Stern--Gerlach apparatus.
\end{abstract}


\section{Introduction}

Let us imagine that we possess an apparatus which performs some 
measurement on certain quantum mechanical system such as spin of
electron. We do not know which measurement is associated with the
device and we would like to find it out.

Obviously, the path to follow here is to
perform set of measurements on various known quantum states and
then estimate the unknown measurement from the collected data.
Such estimation strategy belongs to the broad class of the quantum
reconstruction procedures which attracted considerable attention
recently. The quantum state reconstruction has been widely studied
and now represents a well established technique in many branches
of quantum physics (for a review, see, e.g. \cite{Ulf,Dirk}). The
estimation of the quantum mechanical processes describing
input-output transformations of quantum devices has been discussed
in  \cite{Poyatos97} and
the problem of complete characterization of arbitrary measurement
process has been recently addressed in \cite{LuisPRL}.

Suppose that the  apparatus can respond with $k$ different
measurement outcomes. As is well known from the theory of quantum
measurement \cite{Helstrom},  such device is completely
characterized by $k$ positive operator valued measures (POVM)
$\hat{\Pi}_l$ which govern the measurement statistics,
\begin{equation}
p_{lm}={\rm Tr}[\hat{\Pi}_l \hat{\varrho}_m], \label{plm}
\end{equation}
where $\hat{\varrho}_m$ denotes density matrix of the quantum
state subject to the measurement, $p_{lm}$ denotes the probability
that the apparatus would respond with   outcome  $\hat{\Pi}_l$ to
the quantum state $\hat{\varrho}_m$,
and Tr stands for the trace. The POVMs are positive semi-definite
Hermitian operators,
\begin{equation}
\hat{\Pi}_l \ge 0,
 \label{positivity}
\end{equation}
which decompose the unit operator,
\begin{equation}
\sum_{l=1}^k \hat{\Pi}_l=\hat{I}.
 \label{unity}
\end{equation}
The condition (\ref{positivity}) ensures that $p_{lm}\ge 0$ and
(\ref{unity}) follows from the requirement that total probability
is normalized to unity, $\sum_{l=1}^{k} p_{lm}=1$.

In order to determine the POVMs, one should measure on different
known quantum states $\hat{\varrho}_m$ and then estimate the POVMs
$\hat{\Pi}_l$ from the acquired statistics. Let $F_{lm}$ denote
the total number of detections of $\hat{\Pi}_l$ for the
measurements performed on the quantum state $\hat{\varrho}_m$.
Assuming that the theoretical detection probability $p_{lm}$
(\ref{plm}) can be replaced with relative frequency, we may write
\begin{equation}
{\rm Tr}\,[ \hat{\Pi}_l \hat{\varrho}_m]\equiv \sum_{i,j=1}^N
\Pi_{l,ij}\varrho_{m,ji}=\frac{F_{lm}}{ \sum_{l^\prime=1}^k
F_{l^\prime m} },
 \label{linear}
\end{equation}
where $N$ is the dimension of the Hilbert space on which the
operators $\hat{\Pi}_l$ act.
This establishes a system of linear equations for the unknown
elements of the operators $\Pi_{l,ij}$, which may easily be solved
if  sufficient amount of data is available. This approach is a
direct analogue of linear reconstruction algorithms devised for
quantum state reconstruction. The linear inversion is simple and
straightforward, but it has also one significant disadvantage. The
linear procedure cannot guarantee the required properties of
$\hat{\Pi}_l$, namely the conditions (\ref{positivity}).
Consequently, the linear estimation may lead to unphysical POVMs,
predicting negative probabilities $p_{lm}$ for certain input
quantum states. To avoid such problems, one should resort to more
sophisticated nonlinear reconstruction strategy.

 In this paper we
show that the Maximum-likelihood (ML) estimation is suitable and can be
successfully used for the calibration of the measuring apparatus.
ML estimation has been recently applied to reconstruction of
quantum states \cite{Hradil97,Banaszek99} and
quantum processes  (complete positive maps between density
matrices) \cite{Fiurasek00}. Here we employ it to reconstruct an
unknown {\em quantum measurement} thereby demonstrating again the
remarkable versatility and usefulness of ML estimation.
The generic formalism is developed in Sec. II  and an illustration of
our method is provided in Sec. III, where we estimate a measurement
performed by Stern-Gerlach apparatus.

\section{Maximum-likelihood estimation}

 The estimated operators $\hat{\Pi}_l$  maximize the likelihood
functional
\begin{equation}
{\cal{L}}[\{\hat{\Pi}_l\}]=\prod_{l=1}^k\prod_{m=1}^M
 \left({\rm Tr}[\hat{\Pi}_l\hat{\varrho}_m]\right)^{f_{lm}},
 \label{likelihood}
\end{equation}
where
\begin{equation}
f_{lm}=F_{lm}\left[\sum_{l^\prime=1}^k\sum_{m^\prime=1}^M
F_{l^\prime m^\prime}\right]^{-1}
 \label{f}
\end{equation}
is the relative frequency and $M$ is the number of different
quantum states $\hat{\varrho}_m$ used for the reconstruction.
 The maximum of the likelihood
functional (\ref{likelihood}) has to be found in the sub-space of
physically allowed operators $\hat{\Pi}_l$. We can decompose each
operator $\hat{\Pi}_l$ as
\begin{equation}
\hat{\Pi}_l=\sum_{q=1}^N r_{lq}|\phi_{lq}\rangle \langle
\phi_{lq}|,
\end{equation}
where $r_{lq}\geq 0$ are the eigenvalues of $\hat{\Pi}_l$ and
$|\phi_{lq}\rangle $ are corresponding orthonormal
eigenstates.
The maximum of ${\cal{L}}[\{\hat{\Pi}_l\}]$ can be found from the
extremum conditions.  It is convenient to work with the logarithm
of the original likelihood functional and the constraint
(\ref{unity}) has  to be incorporated by introducing a Hermitian
matrix  of Lagrange multipliers $\lambda_{ij}=\lambda_{ji}^\ast$.
The extremum conditions then read
\begin{eqnarray}
& &
\frac{\partial}{\partial \langle \phi_{lq}| }
\left[\sum_{l=1}^k \sum_{m=1}^M
f_{lm}\ln\left(\sum_{q=1}^N
r_{lq}\langle\phi_{lq}|\hat{\varrho}_{m}|\phi_{lq}\rangle
\right)\right.
\nonumber \\ & & \qquad \quad \left.
-\sum_{l=1}^k \sum_{q=1}^N r_{lq}\langle
\phi_{lq}|\hat{\lambda}|\phi_{lq}\rangle\right]=0.
\label{extremum}
\end{eqnarray}
Thus we immediately find
\begin{equation}
r_{lq}|\phi_{lq}\rangle= \hat{R}_l r_{lq}|\phi_{lq}\rangle,
\label{phiiter}
\end{equation}
where
\begin{equation}
\hat{R}_l=\hat{\lambda}^{-1} \sum_{m=1}^M
\frac{f_{lm}}{p_{lm}}\hat{\varrho}_m \label{Rl}
\end{equation}
and
\begin{equation}
\hat{\lambda}=\sum_{i,j=1}^N \lambda_{ij}|i\rangle\langle j|.
\end{equation}
Let us now multiply (\ref{phiiter}) by $\langle \phi_{lq}| $ from
the right and sum over $q$. Thus we obtain
\begin{equation}
\hat{\Pi}_l= \hat{R}_l \hat{\Pi}_l. \label{Piter}
\end{equation}
On averaging (\ref{Piter}) and its Hermitean conjugate counterpart, we get
\begin{equation}
\hat{\Pi}_l= \frac{1}{2}(\hat{R}_l \hat{\Pi}_l+\hat{\Pi}_l\hat{R}_l^\dagger) .
\label{Piterher}
\end{equation}
The matrix of Lagrange multipliers $\hat{\lambda}$ should  be
determined from the constraint (\ref{unity}). On summing Eq.
(\ref{Piter})  over $l$, we find
\begin{equation}
\hat{\lambda}=\sum_{l=1}^k \sum_{m=1}^M\frac{f_{l m}}{p_{l m}}
\hat{\varrho}_m \hat{\Pi}_l.
 \label{lambdaiter}
\end{equation}
Eqs. (\ref{Piterher}) and (\ref{lambdaiter}) can be conveniently
solved by means of repeated iterations.

If the linear inversion based on Eqs.  (\ref{linear}) provides
physically sensible result, then the ML estimate agrees with this
linear reconstruction. To prove it explicitly, let us assume that
the set of POVMs $\hat{\Pi}_l$ solves the Eqs. (\ref{linear}).
Thus we have
\begin{equation}
p_{lm}=\frac{f_{lm}}{\sum_{l^\prime =1}^k f_{l^\prime m}}.
\end{equation}
On inserting this expression into (\ref{Piter}), we obtain
\begin{equation}
\hat{\Pi}_l={\sum_m}^{\prime}\left(\sum_{l^\prime=1}^k f_{l^\prime
m}\right) \hat{\lambda}^{-1}\hat{\varrho}_m\hat{\Pi}_l.
\label{proof}
\end{equation}
Here the prime indicates that we should sum only over those $m$ with
nonzero $f_{lm}$. However, this restriction may be dropped. If
$f_{lm}=p_{lm}=0$, then $\hat{\varrho}_{m}\hat{\Pi}_l=0$ and the
addition of zero to the right-hand side of (\ref{proof}) changes
nothing. Thus the set of $k$ equations (\ref{proof}) reduces to
\begin{equation}
\hat{\lambda}=\sum_{m=1}^M\sum_{l=1}^k f_{l m} \hat{\varrho}_m
\label{lambda}
\end{equation}
which is the formula for the operator of Lagrange multipliers
valid when the measured data are compatible with some set of POVMs.
Notice that $\hat{\lambda}$ is positive definite.

The differences between linear reconstructions and ML
estimation occur if the experimental data are not compatible with
any physically allowed set of POVMs. The procedure of ML
estimation may be interpreted as a synthesis of information from
mutually incompatible observations. The ML can correctly
handle noisy data and provides reliable estimates in cases when
linear algorithms fail.

Notice that the operators $\hat{R}_l$ contain the inversion of the matrix
$\hat{\lambda}$. The reconstruction is possible only on such
subspace of the total Hilbert space where
the inversion $\hat{\lambda}^{-1}$ exists.
This restriction can easily be understood if we make use of
Eq. (\ref{lambda}).  The experimental data contain only information on
the Hilbert subspace probed by the density matrices
$\hat{\varrho}_{m}$ and the reconstruction of the POVMs must be
restricted to this subspace.

One could complain that it is not certain that the positive
definiteness of POVMs $\hat{\Pi}_l$ is preserved during iterations
based on Eqs. (\ref{Piterher}) and (\ref{lambdaiter}). We can, however,
avoid such complaints by devising an iterative algorithm which exactly
satisfies the constraints (\ref{positivity}) and (\ref{unity})
at each iteration step. We observe that we can formally decompose
the POVMs as
\begin{equation}
\hat{\Pi}_l=\hat{D}_l^\dagger \hat{D}_l,
\end{equation}
where
\begin{equation}
\hat{D}_l=\sum_{q=1}^N  \sqrt{r_{ql}}\,|q\rangle \langle \phi_{lq}|,
\end{equation}
and $|q\rangle$ is some chosen orthonormal basis, $\langle
q|q^\prime\rangle=\delta_{q q^\prime}$.

From the equations (\ref{phiiter}) we can derive
\begin{equation}
\hat{D}_l= \hat{D}_l \hat{R}_l^\dagger .
 \label{Citer}
\end{equation}
The constraint (\ref{unity}) provides formula for the
operator of Lagrange multipliers,
\begin{equation}
\hat{\lambda}^{-1} \hat{G} \hat{\lambda}^{-1}=\hat{I},
\label{lambdaM}
\end{equation}
where $\hat{G}$ is positive operator
\begin{equation}
\hat{G}=\sum_{l=1}^k\sum_{m,m^\prime=1}^M \frac{f_{lm}}{p_{lm}}
\frac{f_{l m^\prime}}{p_{l m^\prime}} \hat{\varrho}_{m^\prime}
\hat{\Pi}_l \hat{\varrho}_m.
\end{equation}
Upon solving (\ref{lambdaM}) we get $\hat{\lambda}=\hat{G}^{1/2}$.
We fix the branch of the square root of $\hat{G}$ by requiring that
$\hat{\lambda}$ should be positive definite operator.
 We can factorize the matrix $\hat{G}$ as
$\hat{G}=\hat{U}^\dagger \hat{\Lambda} \hat{U}$
where $\hat{U}$ is unitary matrix and $\hat{\Lambda}$ is diagonal
matrix containing eigenvalues of $\hat{G}$. We define
\begin{equation}
 \hat{\Lambda}^{1/2}={\rm
 diag}(\sqrt{\Lambda_{11}},\ldots,\sqrt{\Lambda_{NN}})
\end{equation}
and we can write
\begin{equation}
\hat{\lambda}=\hat{U}^\dagger \hat{\Lambda}^{1/2} \hat{U}.
 \label{lambdasqrt}
\end{equation}

The advantage of the iterative procedure based on (\ref{Citer})
and (\ref{lambdasqrt}) is that both conditions (\ref{positivity})
and (\ref{unity}) are exactly fulfilled at each iteration step.
The disadvantage of this approach is the greater numerical
complexity in comparison to iterations based on (\ref{Piter}) and
(\ref{lambdaiter}), because we must calculate the eigenvalues of
the matrix $\hat{G}$ at each iteration step.

The determination of the quantum measurement  can simplify considerably
if we have some {\em a-priori} information about the apparatus.
For example, if we know that we deal with a photodetector, then
we have to estimate only a single parameter,
the absolute photodetection effeciency $\eta$ \cite{DAriano00}.
Here we briefly consider a broader class of phase-insensitive detectors
which are sensitive only to the number of photons in a single mode of
electromagnetic field. The POVMs describing
phase-insensitive detector are all diagonal in the Fock basis,
\[
\hat{\Pi}_l=\sum_n r_{ln} |n\rangle \langle n|
\]
and the ML estimation reduces to the determination of the
eigenvalues $r_{ln}\geq 0$. The extremum Eqs. (\ref{Piterher}) and
(\ref{lambdaiter}) simplify to
\begin{eqnarray}
r_{ln}&=&\frac{ r_{ln}}{\lambda_n}\sum_{m=1}^M
\frac{f_{lm}}{p_{lm}}\varrho_{m,nn}, \nonumber \\
 \lambda_n&=&
 \sum_{m=1}^M \sum_l \frac{f_{l m}}{p_{lm}}\varrho_{m,nn} r_{l n},
\nonumber \\
 p_{lm}&=&\sum_n \varrho_{m,nn} r_{ln}.
  \end{eqnarray}

 Instead of solving the extremum equations,
 one may directly search for the maximum of
${\cal{L}}[\{\hat{\Pi}_l\}]$ with the help of downhill-simplex
algorithm \cite{Banaszek99}. To implement this algorithm
successfully, it is necessary to use a minimal parametrization. If
we deal with $N$ level system, then each $\hat{\Pi}_l$ is
parametrized by $N^2$ real numbers. Since the constraint
(\ref{unity}) allows us to determine one POVM in terms of
remaining $k-1$ ones, the number of independent real parameters
reads $N^2(k-1)$. Furthermore we may take advantage of the
Cholesky decomposition,
\begin{equation}
\hat{\Pi}_l= \hat{C}_l^\dagger \hat{C}_l,
 \label{Cholesky}
\end{equation}
where $\hat{C}_l$ is lower triangular matrix with real elements on
its main diagonal. The parametrization (\ref{Cholesky}) is used
for the first $k-1$ operators, and the last one is calculated form
(\ref{unity}),
\begin{equation}
\hat{\Pi}_k=\hat{I}-\sum_{l=1}^{k-1}\hat{C}_l^\dagger \hat{C}_l,
\label{Pk}
\end{equation}
thus achieving minimal parametrization. For each parameter set,
where ${\cal{L}}[\{\hat{\Pi}_l\}]$ is evaluated, one has to check
whether (\ref{Pk}) is positive semi-definite. If this does not
hold, then one sets ${\cal{L}}[\{\hat{\Pi}_l\}]=0$, thus
restricting the numerical search of the maximum to the domain of
physically allowed operators. This domain is a finite volume
subspace of a $N^2(k-1)$ dimensional space.

\section{Stern-Gerlach apparatus}

In this section we illustrate the developed formalism by means of
numerical simulations for Stern-Gerlach apparatus measuring
a spin-$1$  particle. We choose the three
eigenstates of the operator of $z$-component of the spin as the basis states,
$\hat{\sigma}_z|s_z\rangle=s_z|s_z\rangle$, $s_z=-1,0,1$. In this
basis, the matrix representation of the spin operators reads
\[
\hat{\sigma}_x= \frac{1}{\sqrt{2}}
 \left(
\begin{array}{ccc}
 0 & 1 & 0 \\ 1 & 0 & 1 \\ 0 & 1 & 0
\end{array}
\right), \qquad
 \hat{\sigma}_y= \frac{1}{\sqrt{2}}
 \left(
\begin{array}{ccc}
 0 & -i & 0 \\ i & 0 & -i \\ 0 & i & 0
\end{array}
\right),
\]
\[
\hat{\sigma}_z=
 \left(
\begin{array}{ccc}
 1 & 0 & 0 \\ 0 & 0 & 0 \\ 0 & 0 & -1
\end{array}
\right).
\]

\begin{figure}[t]
\centerline{\psfig{figure=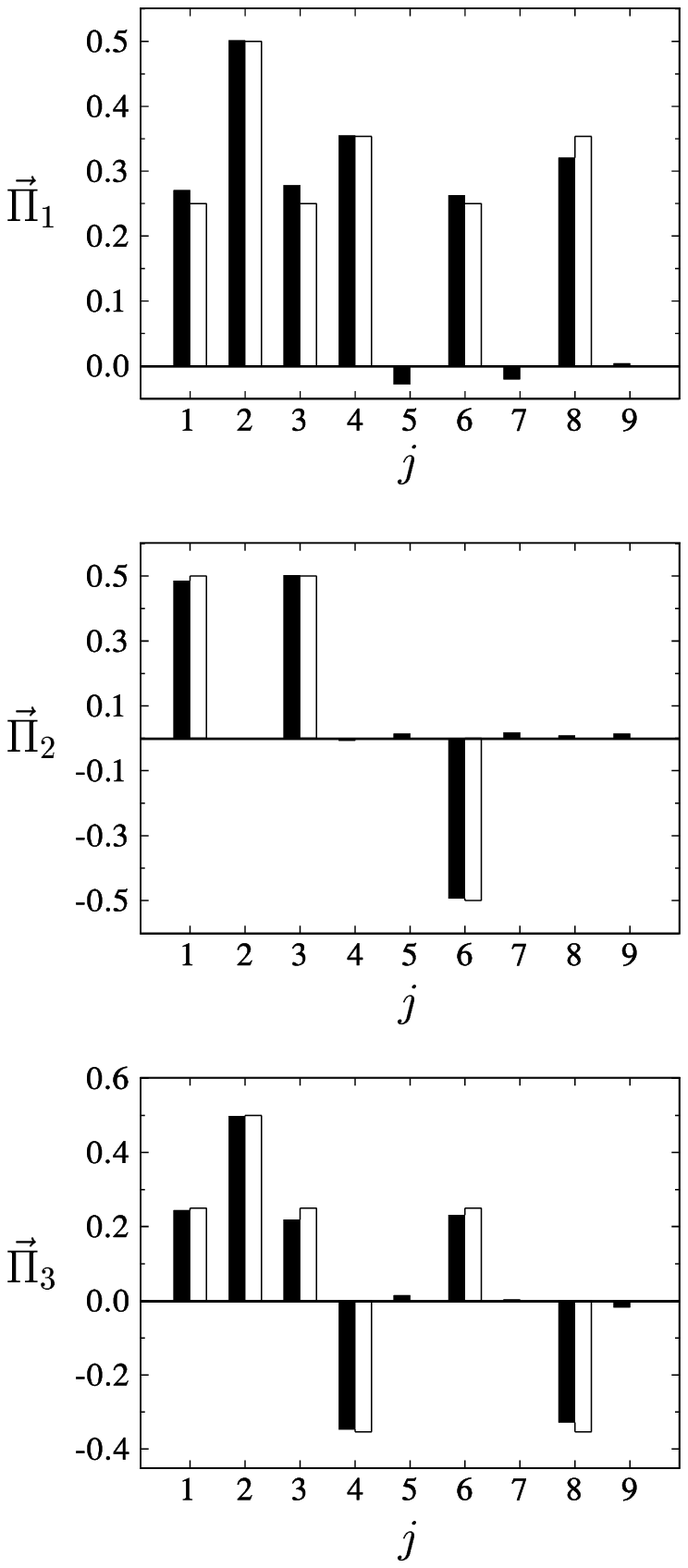,width=0.64\linewidth}}
\small FIG. 1:
Reconstructed POVMs $\hat{\Pi}_l$ (solid
bars) and the exact POVMs (hollow bars). The figure displays
elements of the vectors $\vec{\Pi}_l$ defined in Eq.
(\ref{Pivector}). In the simulation, we made ${\cal{N}}=30$
measurements on each of $12$ different quantum states,
which represents altogether $360$ measurements.
\end{figure}

In our numerical simulations, we assume $12$ different pure quantum
states: three eigenstates of $\hat{\sigma}_z$, $|-1_z\rangle$,
$|0_z\rangle$, and $|1_z\rangle$, and nine superposition states
$2^{-1/2}(|-1_z\rangle+e^{i\psi_j}|0_z\rangle),$
$2^{-1/2}(|0_z\rangle+e^{i\psi_j}|1_z\rangle),$ and
$2^{-1/2}(|-1_z\rangle+e^{i\psi_j}|1_z\rangle),$
where $j=1,2,3$ and $\psi_1=0$, $\psi_2=\pi/2$, and $\psi_3=\pi$.
 The measurement
on each state is performed $\cal{N}$ times thus the total number
of measurements is $12\cal{N}$. We have performed Monte Carlo
simulations of the measurement and then we have reconstructed the
three POVMs $\hat{\Pi}_l$ characterizing given apparatus. In the
simulations, the device measures the projection of the spin along
$x$ axis, and the true POVMs  are projectors
\begin{equation}
\hat{\Pi}_1=|1_x\rangle\langle 1_x|,\quad
 \hat{\Pi}_2=|0_x\rangle\langle 0_x|,\quad
\hat{\Pi}_3=|-1_x\rangle\langle -1_x|, \label{truePOVM}
\end{equation}
$\hat{\sigma}_x|s_x\rangle = s_x |s_x\rangle$.
The elements of the POVMs can be conveniently collected in a real
vector
\begin{eqnarray}
 \vec{\Pi}_l&=&(\Pi_{l,11},\, \Pi_{l,22},\,
 \Pi_{l,33}, \, {\rm Re} \,\Pi_{l,12}, \, {\rm Im}\, \Pi_{l,12},
 \nonumber \\
 && \,\,{\rm Re}\, \Pi_{l,13},\,
 {\rm Im} \, \Pi_{l,13}, \, {\rm Re} \,\Pi_{l,23}, \, {\rm Im}\, \Pi_{l,23}
 ).
 \label{Pivector}
\end{eqnarray}
The estimated POVMs as well as the 'true' POVMs (\ref{truePOVM})
are plotted in Fig. 1. The estimate has been obtained by iteratively
solving the extremum equations and
both sets (\ref{Piterher}), (\ref{lambdaiter}) and (\ref{Citer}),
(\ref{lambdasqrt}) provide identical results.
The reconstructed operators are in good agreement with the exact ones.
Equally important is the fact that
the estimated operators $\hat{\Pi}_l$ meet the constraints
(\ref{positivity}) and (\ref{unity}).

In summary, we have shown how to reconstruct a generic quantum
measurement with the use of  Maximum-likelihood principle. Our method
guarantees that the estimated POVMs, which fully describe the
measuring apparatus, meet all the required positivity and completness
constraints. The numerical feasibility of our technique has been
illustrated by means of numerical simulations for Stern-Gerlach apparatus.

\acknowledgements
I would like to thank Z. Hradil, M. Je\v{z}ek, and J.
\v{R}eh\'{a}\v{c}ek for stimulating
discussions. This work was supported by
Research Project CEZ: J14/98: 153100009 ``Wave and Particle Optics''
of the Czech Ministry of Education.

\end{document}